# On-chip wavelength division multiplexing by angled multimode interferometer fabricated on erbium-doped thin film lithium niobate on insulator

Jinli Han[1,2], Rui Bao[1,2], Rongbo Wu[2], Zhaoxiang Liu[2], Zhe Wang[2], Chao Sun[1,2], Zhihao Zhang[2], Mengqi Li[2], Zhiwei Fang[2], Min Wang[2], Haisu Zhang[2,\*], and Ya Cheng[1,2,3,4,5,6,\*]

[1]State Key Laboratory of Precision Spectroscopy, East China Normal University, Shanghai 200062, China.
[2]The Extreme Optoelectromechanics Laboratory (XXL), School of Physics and Electronic Sciences, East China Normal University, Shanghai 200241, China.
[3]State Key Laboratory of High Field Laser Physics and CAS Center for Excellence in Ultra-Intense Laser Science, Shanghai Institute of Optics and Fine Mechanics (SIOM), Chinese Academy of Sciences (CAS), Shanghai 201800, China.
[4]Shanghai Research Center for Quantum Sciences, Shanghai 201315, China.
[5]Collaborative Innovation Center of Extreme Optics, Shanxi University, Taiyuan 030006, China.
[6]Collaborative Innovation Center of Light Manipulations and Applications, Shandong Normal University, Jinan 250358, China.
\*Correspondence: Haisu Zhang (hszhang@phy.ecnu.edu.cn), Ya Cheng (ya.cheng@siom.ac.cn).

## Abstract

Photonic integrated circuits based on erbium doped thin film lithium niobate on insulator has attracted broad interests with insofar various waveguide amplifiers and microlasers demonstrated. Wideband operation facilitated by the broadband absorption and emission of erbium ions necessitates the functional integration of wavelength filter and multiplexer on the same chip. Here a low-loss wavelength division multiplexer at the resonant pumping and emission wavelengths (~1480 nm and 1530~1560 nm) of erbium ions based on angled multimode interferometer, is realized in the erbium doped thin film lithium niobate on insulator fabricated by the photolithography assisted chemomechanical etching technique. The minimum on-chip insertion losses of the fabricated device are <0.7 dB for both wavelength ranges, and a 3-dB bandwidth of >20 nm is measured at the telecom C-band. Besides, direct visualization of the multimode interference pattern by the visible upconversion fluorescence of erbium ions compares well with the simulated light propagation in the multimode interferometer. Spectral tuning of the wavelength division multiplexer by structural design is also demonstrated and discussed.

# Introduction

Stimulated by the excellent nonlinear optical and electro/acousto-optic properties of lithium niobate, the broad optical gain by excited erbium ions in the telecom C-band and the maturing advancement of microfabrication technique[1-3], photonic integrated circuits (PIC) on the erbium doped thin film lithium niobate on insulator (TFLN) platform has evolved rapidly with an abundant of erbium-doped TFLN devices including waveguide amplifiers[4-8], micro-resonator lasers[9-15], and optical quantum memories[16-18]. Compared to the low refractive-index-contrast lithium niobate waveguide fabricated by titanium-ion diffusion or proton exchange in bulk crystals[19-21], high-contrast waveguide by lithographically patterning lithium niobate thin film of submicron thickness allows tight optical confinement and thus efficient light-matter interaction and dense integration. Therefore, the demonstrated erbium-doped TFLN devices excel in power consumption, footprints and scalability when compared to their bulk counterparts[4-21].

To achieve high spectral efficiency in the telecom C-band, wavelength filtering and multiplexing in erbium-doped TFLN devices is desired despite the various demonstration of such functionalities in passive TFLN platforms[22-24]. Due to the low efficiency of selective doping of TFLN by erbium ion implantation or diffusion, direct stitching of active and passive TFLN wafer has been employed for hybrid integration, though the design versatility and fabrication throughput is still limited in the stitched active/passive wafer[25]. Alternatively, direct integration of the functional components in erbium-doped TFLN is viable for compact multifunctional PICs such as the electro-optically tunable microlasers and Sagnac-loop reflector based Fabry-Perot resonator lasers demonstrated recently[14,15].

In this work, a low-loss two-channel wavelength division multiplexer (WDM) at the resonant pumping and emission wavelengths (~1480 nm and 1530~1560 nm) of erbium ions based on angled multimode interferometer (MMI) is realized in the erbium-doped TFLN platform. The designed structure featuring a dispersive self-imaging of the input wavelengths at varying lengths is fabricated by the photolithography assisted chemomechanical etching (PLACE) technique[26,27]. The minimum on-chip insertion losses of the fabricated device are <0.7 dB for both wavelength ranges, and a 3-dB bandwidth of >20 nm is measured at the telecom C-band. Thanks to the upconversion fluorescence of erbium ions in the WDM device, the dynamic multimode interference pattern is clearly observed showing high consistence with simulation. Spectral tuning of the WDM device is also demonstrated and discussed.

# Results

Design of angled multimode interferometer

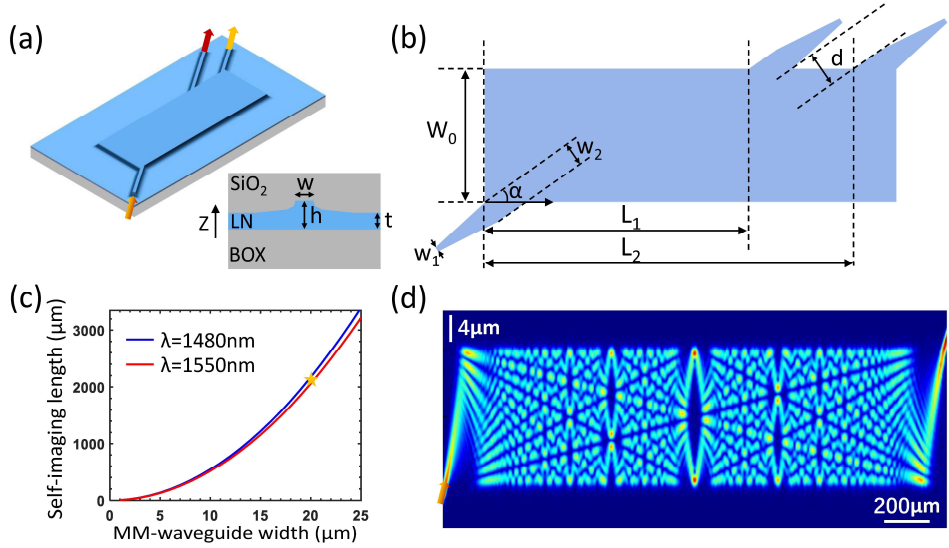

Fig. 1 Design of the angled-MMI based WDM device. (a) the schematic of the device, (b) the relevant parameters defining the angled-MMI structure, (c) the self-imaging length vs the MMI-width at two different wavelengths, (d) the simulated light propagation profile in the angled-MMI device. The inset in (a) is the cross-section profile of the TFLN waveguide.

The schematic of the angled-MMI based WDM device is shown in Figs. 1(a) and 1(b), where the input and output waveguides are inclined at an angle $\alpha$ with respect to the central multimode waveguide facilitating a dispersive self-imaging of the input modes at different wavelengths into separated locations. The MMI self-imaging length $L_i$ at the wavelength $\lambda_i$ is well described by the effective index $n_{\text{eff}}$ of the fundamental mode and the width $W_0$ of the multimode waveguide as[28]:

$$L_i = \frac{4n_{\text{eff}}W_0^2}{\lambda_i} \qquad (1)$$

The effective index of the fundamental mode of the multimode waveguide is calculated with the waveguide cross section shown in the inset of Fig. 1(a), where a ridge profile with a remaining slab layer is formed in the lithium niobate thin film with the overlaid silica cladding. The erbium-doped Z-cut TFLN with the LN-height of h=500 nm is employed in this work. Then using Eq. (1) the self-imaging lengths at $\lambda_1$=1550 nm ($L_1$) and $\lambda_2$=1480 nm ($L_2$) are plotted versus the width of the multimode waveguide in Fig. 1(c). For a trade-off between the device footprint and the fabrication resolution, the MMI width is selected at $W_0 = 20$ μm and the corresponding self-imaging lengths are ~2.2 mm with the separation between the two wavelengths to be $\Delta L = L_2 - L_1 \approx 120$ μm labelled by the yellow star in Fig. 1(c). For a small inclining angle $\alpha$=6°, the vertical distance between the two output waveguides is $d = \Delta L \sin \alpha - w_2 \approx 9.5$ μm, where $w_2 = 3$ μm is the width of the input and output waveguides projected on the MMI. The input and output waveguides are further tapered to $w_1 = 1$ μm for close to single-mode-excitation. The simulated light propagation profile for the inclined incidence into the MMI is shown in Fig. 1(d).

The sensitivity of the transmission loss of the designed WDM device on the structural parameters are further simulated by the eigen-mode-expansion method. When the etching depth of the TFLN is tuning from 210 nm to 270 nm (with corresponding slab thickness from $t$=290 nm to $t$=230 nm), the self-imaging length at $\lambda_1$=1550 nm is shifted by the value from 40 μm to -32 μm compared to the self-imaging length $L_1$ at the optimized etching depth of 240 nm ($t$=260 nm) as shown in Fig. 2(a), while for the fixed $L_1$ the peak transmission wavelength is changing from 1575 nm to 1530 nm shown in Fig. 2(b). Similar results are also obtained by tuning the MMI width around $W_0 = 20$μm from -90 nm to 90 nm shown in Figs. 2(c) and 2(d), where the moving of the self-imaging length at $\lambda_1$=1550 nm as well as the shifting of the central wavelength at fixed $L_1$ are obtained. Moreover, at the optimized MMI width of $W_0 = 20$ μm and etching depth of 240 nm, the peak transmission wavelengths can be adjusted from 1520 nm to 1560 nm with the 3-dB bandwidth of 15 nm and the maximum cross-talks below -20 dB by changing the output waveguide position as shown in Figs. 2(e) and 2 (f). It should be noted that the transmission losses at the peak wavelengths are all above -1dB denoted as grey dashed lines in Figs. 2(a-f) considering the fabrication fluctuation of the angled-MMI, and the designed two-channel WDM device is aimed for in-band pumping of erbium doped waveguide amplifiers and lasers which have high tolerance for the signal transmission bands.

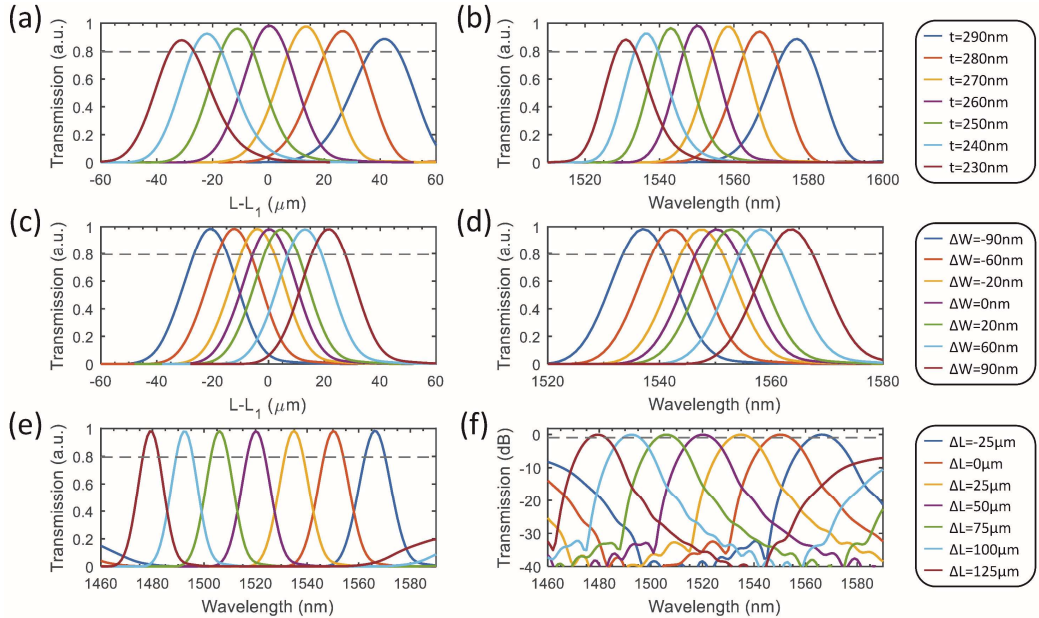

Fig. 2 MMI transmission versus structure parameters. The transmission at $\lambda_1$=1550 nm vs output waveguide positions and the transmission spectra at fixed output waveguide position for (a-b) slab thickness variation and (c-d) MMI width variation, (e-f) the transmission spectra at fixed slab thickness and MMI width but different output waveguide positions. The same legends apply for (a) and (b), (c) and (d), (e) and (f), respectively.

## Device fabrication and characterization

The designed angled-MMI based WDM device is fabricated by the PLACE technique mainly consisting of mask patterning by femtosecond laser direct writing and film etching by chemomechanical polishing. The 500 nm-thick Z-cut TFLN wafer produced by the "smart-cut" technology (NanoLN, Jinan Jingzheng) from the congruent erbium-doped lithium niobate bulk crystal with the doping concentration of 0.5 mol% is adopted for device fabrication. A thin layer (1.5 μm) of $SiO_2$ film is then deposited by the inductively coupled plasma chemical vapor deposition (IPCVD) at low temperature of 80 °C on top of the fabricated structure. The fabrication details by PLACE can be found in previous works[26]. The designed footprint is shown in Fig. 3(a), and the optical microscope image of the fabricated on-chip WDM device in shown in Fig. 3(b). The scanning electron microscope (SEM) images for the input waveguide, the multimode waveguide, and the output waveguides are further shown in the enlarged views (i-iii) of Fig. 3(b). The cross-sectional image of the fabricated TFLN waveguide before the silica cladding deposition is shown in Fig. 3(c), where the ridge profile can be clearly identified. The simulated fundamental transverse-electric mode ($TE_{00}$) of the waveguide with the top width of 1 μm and silica cladding is depicted in Fig. 3(d), and the first higher-order mode ($TE_{01}$) supported by the waveguide is also shown. Due to the slant sidewall of the waveguide fabricated by the PLACE technique, higher-order modes are usually allowed to propagate which can be suppressed by careful alignment of the coupling fiber on the input facet for close to single-mode-excitation.

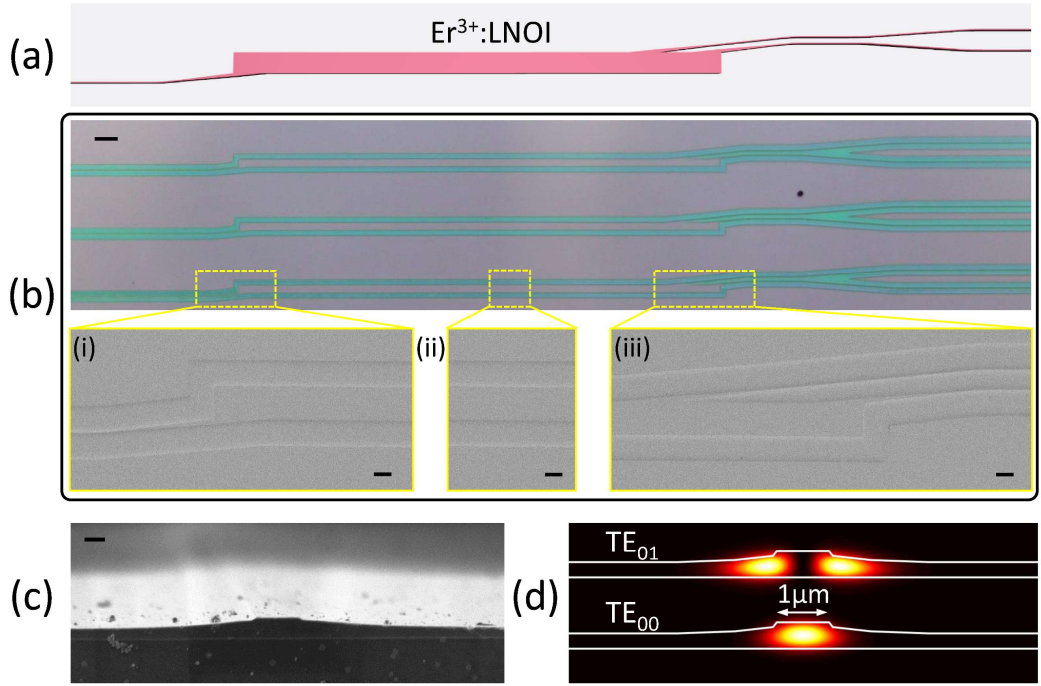

Fig. 3 Microscope images of the fabricated WDM device. (a) the designed footprint, (b) the optical microscope image of the on-chip WDM with the inset showing the enlarged SEM-views, (c) the SEM-image of the waveguide cross section, (d) the simulated optical mode intensity profiles. The scale bars in (b), insets (i-iii) of (b) and (c) are 50 μm, 10 μm, and 500 nm, respectively.

To characterize the fabricated on-chip WDM device, a fiber coupled high power laser diodes at the wavelength of 1480 nm is first connected to lensed fibers with the cleaving angle of 99° for butt-coupling with the input waveguides of the WDM device. A fiber-based polarization controller is used to tune the input polarization state for fundamental TE mode excitation. Thanks to the visible upconversion fluorescence from erbium ions excited by the high power 1480 nm pump laser, the in-situ light propagation profile within the MMI can be easily inferred from the top-view imaging as shown in Fig. 4(a). The captured fluorescence images are shown in Fig. 4(b), where the incident laser from the left side and a multiple of fractional images of the input mode and the full mirror image are observed[28]. The simulated light propagation profiles are also shown in Fig. 4(b) for comparison. A remarkable consistence between the measured fluorescence images and the simulated multimode interference patterns can be clearly noticed, confirming the desired function of the fabricated angled-MMI as a self-imaging device. Besides, the fluorescence images around the output position of the angled-MMI are shown for three different output lengths in Fig. 4(c), respectively. It can be clearly seen that when the input light self-images before/after the output waveguide position (see the images (i) and (iii) in Fig. 4(c)), multiple reflection and scattering on the MMI sidewalls will induce high transmission losses. High transmission is obtained when the output waveguide is located around the self-imaging length with a tolerance range of $ΔL=±5$ μm as found in the experiment.

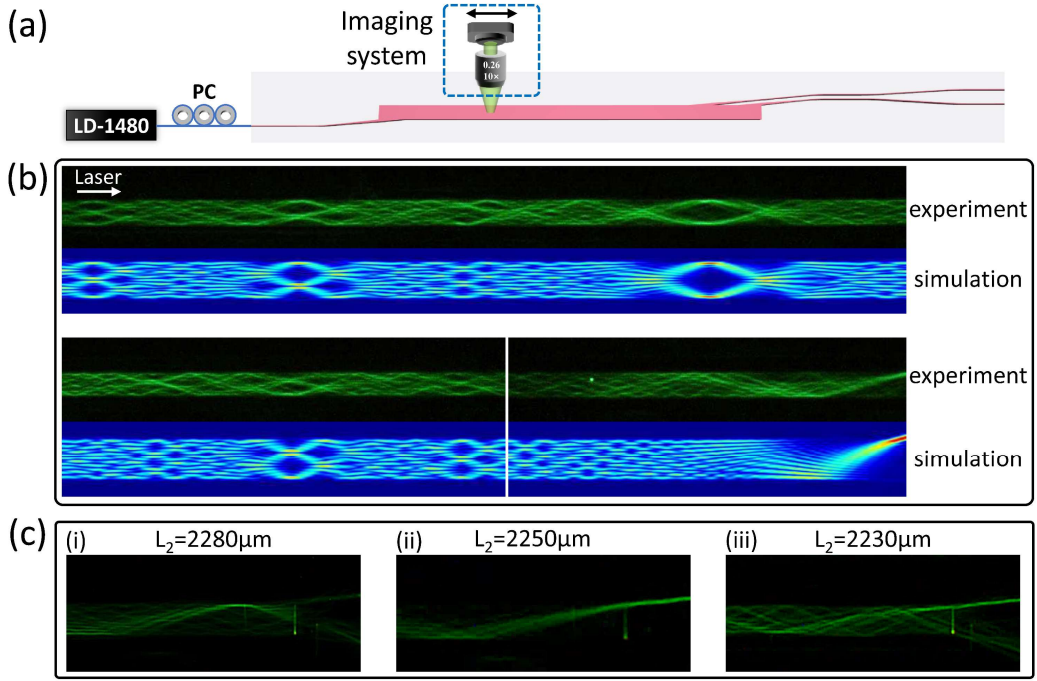

Fig. 4 Fluorescence imaging of the on-chip WDM device. (a) the experimental schematic, (b) the measured fluorescence microscope images with incident pump

laser from the left side as well as the simulated multimode interference profiles. (c) the fluorescence images at the output region for three output waveguide positions (i) $L_2$=2280 μm, (ii) $L_2$=2250 μm and (iii) $L_2$=2230 μm.

The insertion losses of the fabricated on-chip WDM device are further measured by injecting the 1480 nm pump laser and the C-band tunable continuous-wave laser signals into the common input port and monitoring the output powers at the two output ports by a calibrated power meter. The experimental setup is shown in Fig. 5(a), where the microscope imaging of the output modes by an InGaAs infrared camera is also depicted for clarity. The on-chip transmission losses are then obtained by comparing the insertion losses of the WDM device with the straight waveguide of the same length, in such way the fiber-chip facet coupling losses as well as the waveguide propagation losses including erbium absorption loss are removed with the remaining losses exclusively arising from the on-chip wavelength multiplexing.

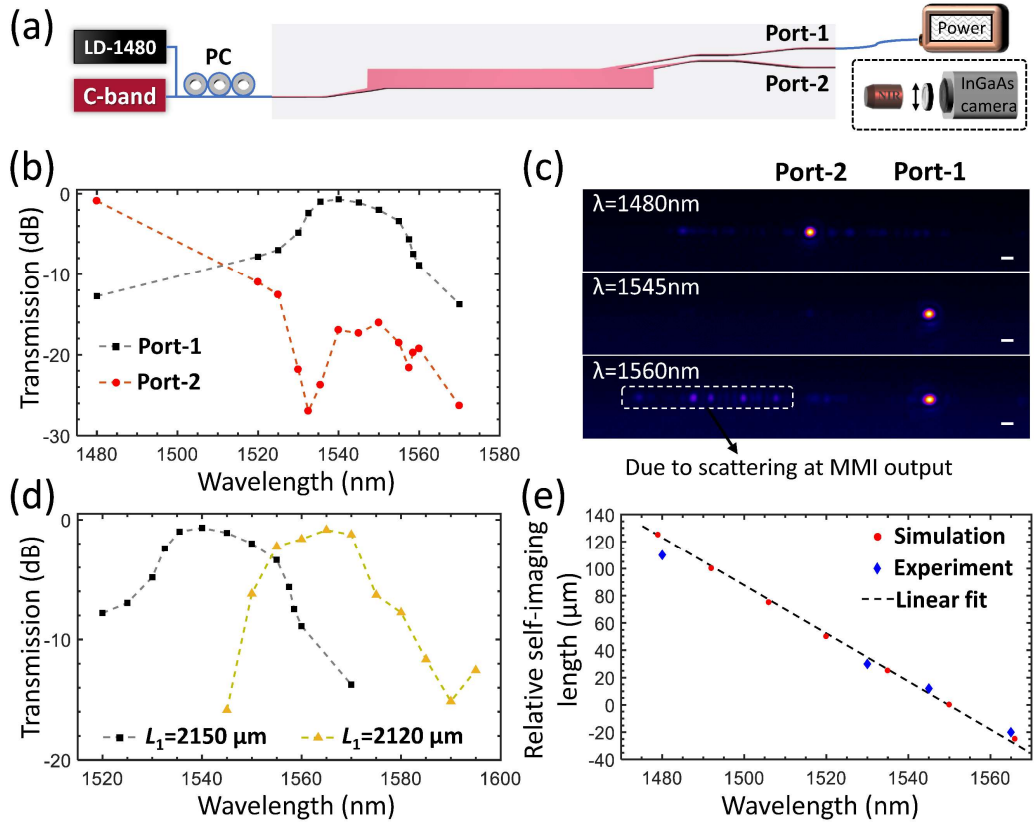

Fig. 5 Transmission loss measurement of the WDM device. (a) the experimental setup, (b) the transmission spectral response of Port-1 and Port-2, (c) the output mode profiles both ports, (d) the transmission band with respect to Port-1 position ($L_1$), (e) the simulated and measured relative self-imaging lengths at different wavelengths. The scale bars in (c) are 5 μm.

The measured and calibrated transmission losses of the on-chip WDM device are plotted in Fig. 5(b). The transmission from Port-1 is maximized at $\lambda_1$=1540 nm with the minimum loss of 0.68 dB and the transmission losses are smaller than 3 dB for the wavelength range from 1532 nm to 1555 nm. The 1480 nm pump laser transmission loss from Port-2 is around 0.52 dB and the transmission losses from Port-2 are larger than 18 dB for the 3-dB transmission band of Port-1 (1532 nm-1555 nm). The output mode profiles from Port-1 and Port-2 are captured in the far field as shown in Fig. 5(c), where the output modes are located at Port-1 and Port-2 when the incident light wavelengths are 1545 nm and 1480 nm, respectively. The output mode at the incident wavelength of 1560 nm is largely degraded by the strong scattering from the MMI sidewalls induced by the self-imaged light before the output waveguide position (the longer wavelength signal will self-image earlier). Meanwhile, the transmission band of Port-1 can also be adjusted by changing the output waveguide position as shown in Fig. 5(d), where the peak transmission wavelength of Port-1 is shifted from $\lambda_1$=1540 nm to $\lambda_1$=1565 nm with comparable transmission loss and 3-dB bandwidth when the output waveguide position is moved from $L_1$=2150 μm to $L_1$=2120 μm. The simulated wavelength-dependent self-imaging lengths are plotted as circles in Fig. 5(e), with a linear fitting by the dashed lines. The experimentally measured peak transmission wavelengths at four different self-imaging lengths are also plotted as diamonds in Fig. 5(e), which fall well around the dashed lines fitted from simulation.

## Discussion

On-chip wavelength multiplexing and demultiplexing has attracted broad interests in integrated photonics due to their broad applications in optical communication, optical interconnect, remote sensing and ranging. A diversity of design protocols have been demonstrated in various platforms, including arrayed waveguide gratings (AWG), tunable microring filters, cascaded Mach-Zehnder interferometer (MZI), MZI with bent directional couplers, cascaded multimode waveguide gratings (MWG) and angled multimode interferometers (MMI)[29-34]. Among the various designs the angled MMI have been considered as an efficient multichannel device for coarse wavelength division multiplexing (CWDM) in the C-band as well as the O-band[23,33,34]. The current work employing the angled-MMI as a two-channel wavelength multiplexer is mainly aimed for in-band pumping of the on-chip erbium doped waveguide amplifiers and lasers by the 1480 nm pump laser as well as the injection and extraction of the amplified C-band signals emitted by excited erbium ions. The adopted fabrication technique by PLACE is also featured with low-loss planar waveguides with smooth sidewalls of sub-nanometer roughness comparable with the surface-tension, though with a compromised resolution for fine structure pattering and etching[26,27]. The designed and fabricated two-channel WDM device with low transmission losses proves its suitability for the PLACE fabrication technique.

In conclusion, a low-loss wavelength division multiplexer (WDM) at the resonant pumping and emission wavelengths (~1480 nm and 1530~1560 nm) of erbium ions based on angled multimode interferometer (MMI) is designed and fabricated in the erbium-doped TFLN platform. The minimum on-chip insertion losses of the fabricated device are <0.7 dB for both wavelength ranges, and a 3-dB bandwidth of >20 nm is measured at the telecom C-band. Upconversion fluorescence of erbium ions in the WDM device also visualize the dynamic multimode interference pattern which is in full consistence with simulation. The demonstrated on-chip 1480/1550 WDM will find great use in integrated waveguide amplifiers and lasers based on the erbium-doped thin film lithium niobate on insulator platform.